\documentstyle[12pt,psfig,epsfig]{article}

\newcommand{\be}{\begin{eqnarray}}
\newcommand{\ee}{\end{eqnarray}}

\def\nue{{\nu_e}}
\def\anue{{\bar\nu_e}}
\def\numu{{\nu_{\mu}}}
\def\anumu{\bar\nu_{\mu}}
\def\nutau{{\nu_{\tau}}}
\def\anutau{{\bar\nu_{\tau}}}

\newcommand{\dm}{\mbox{$\Delta{m}^{2}$~}}

\newcommand{\chr}{\mbox{$\breve{\rm C}$erenkov~}}

\begin{document}
\begin{center}
{\large \bf Neutrinos from Supernovae }
\vspace{.5in}

{\bf Sandhya Choubey$^a$\footnote{Now at, 
INFN Sezione di Trieste and
Scuola Internazionale Superiore di Studi Avazanti,
I-34014,
Trieste, Italy.}
and Kamales Kar$^b$}
\vskip .5cm

$^a${\it Department of Physics and Astronomy, University of Southampton, \\
Highfield, Southampton S017 1BJ, UK}\\
$^b${\it Theory Group, Saha Institute of Nuclear Physics,\\
Kolkata 700 064, India}

\vskip 0.5in

\end{center}

\begin{abstract}
In this review, the effect of flavor oscillations on the neutrinos
released during supernova explosion after core collapse is
described. In some scenarios there are large enhancement of the number 
of events compared to the no oscillation case. Various other features
associated with supernova neutrinos are also discussed.
\end{abstract}

\section{Introduction}

February 23, 1987 saw the birth of a new era in astrophysics -- 
extra--solar system
neutrino astronomy. The supernova explosion in the Large Magellanic Cloud (LMC)
at a distance of about 50 kpc was not only the closest visual supernova
since Kepler but was also the source of neutrinos detected at the terrestrial
detectors of Kamiokande (KII) and IMB giving rise to 11 and 8 events respectively.
 The next few years saw great excitement in this field. Astrophysics interacted
with particle physics intimately. From the number and the energy distribution
of the observed neutrinos one tried to extract information about the stellar
core and check them with model predictions. On the other hand these
neutrinos also gave particle physics constraints on neutrino properties.
In the last few years interest in this area got rejuvenated by the finding
that neutrinos do have non-zero mass and the flavors do mix when they travel.
This conclusion was reached through the analysis of the atmospheric neutrinos
detected at the Superkamioka (SK) along with their zenith angle dependence
and the observation of the deficit of detected solar neutrinos by the Chlorine
and Gallium radiochemical 
detectors and at SK and Sudbury Neutrino Observatory (SNO) through
electron scattering and charged/neutral current dissociation of heavy water
respectively. The recent results announced by the KamLAND reactor experiment
gives for the first time conclusive evidence for neutrino oscillation 
using a terrestrial neutrino source and confirms the Large 
Mixing Angle (LMA) solution to the solar neutrino problem.
Thus the present day interest in supernova neutrinos lies around
the question: if you have a galactic supernova event today what would be the
number of events and their time and energy distributions in the large number of
neutrino detectors in operation. The other related question is whether one can
get a signature of neutrino oscillation mechanism from the observed data
and also how other neutrino properties get constrained. Information about the
mechanisms of the supernova explosion is also an area of huge  
interest. In this review we survey some of these issues. In section 2 we
give a brief overview of the physics of type II supernovae and 
the emission of neutrinos from them. Section 3 introduces
the subject of neutrino oscillation and the impact of vacuum and 
matter enhanced
oscillation on the supernova neutrinos from the core. Section 4 describes the
expected number of events in the terrestrial detectors for the different mass
and mixing scenarios. Finally section 5 briefly states the other connected 
issues of suernova neutrino detection.

\section{Type II Supernovae and Neutrino Emission}

Stars of masses larger than $8 M_\odot  $ after burning for millions of years
collapse when the nuclear reactions in the core stop with matter consisting  
mostly of $^{56}Fe$ like nulcei. This collapse proceeds very fast (timescale
 of the order of tens of milliseconds) and stops in the central region when 
its density goes beyond the nuclear matter density with
a strong shock starting to travel outward \cite{Bethe:1990}. This shock wave,
eventually hitting the outer mantle in a few seconds and supplying the 
explosion
energy of a few times $10^{51}$ ergs, is believed to be the cause of type II 
supernova explosion. During this process, the binding energy released comes
out almost completely as neutrinos and antineutrinos of three different
flavors (e, $\mu$ and $\tau$) in the ``cooling phase'' with the total energy
release of the order of $10^{53}$ ergs. Let us discuss the emission of the
neutrinos in some more detail. Firstly during the early stage of the collapse
(densities less than $10^{12}$ g/cc) neutrinos are produced through 
neutronisation
\be
     e^- + (N,Z) \rightarrow  (N+1,Z-1) + \nue     
\ee
\be
     e^- + p     \rightarrow   n + \nue 
\ee
  where only $\nue$ (not $\anue$) are produced. 
At lower densities 
these neutrinos 
have mean free path much larger than the core radius and hence
escape. But the total energy of these neutronisation neutrinos 
is much smaller than that
in the cooling phase. Even then it is possible to detect them for nearby 
galactic
supernovae at distances within 1 kpc \cite{Sutaria:1997}. These neutrinos can 
give information about the temperature and composition of the core.

The main neutrino emission is during the cooling 
phase where the thermal $\nu$/$\bar\nu$ are 
produced through pair production and other processes \cite{Burrows:1990}. 
Out of these
$\numu$, $\nutau$, $\anumu$ and $\anutau$, called collectively as $\nu_x$,
 interact with matter only through neutral current whereas $\nue$ and $\anue$
have both charged current and neutral current interaction with matter. As the
matter is neutron-rich the $\nue$'s interact more with matter than the 
$\anue$'s. These neutrinos deep
inside the core are in equilibrium with the ambient matter 
and their energy distributions are close to 
Fermi-Dirac as seen through simulations and through the analysis of 1987A
neutrinos \cite{Schramm:1990}. As the stellar core has a strong density 
gradient, electron type
neutrinos can stay in equilibrium upto larger radius and so the $\nue$ 
``neutrinosphere'' has the largest radius and smallest temperature. In this 
article
we shall assume that the three types of neutrino gas have Fermi-Dirac 
distributions with temeratures 11, 16 and 25 MeV for $\nue$, $\anue$ and 
$\nu_x$ respectively.

An important role played by neutrinos in type II supenovae is in the  
process of ``delayed neutrino heating'' \cite{Wilson:1985}. In almost all
simulations for large mass stars one sees that the shock wave moving outward
fast loses energy in dissociating the nuclei in the overlying matter and
soon becomes an accretion shock. This shock gets revitalised over the much 
longer timescale of seconds through the absorption of a small fraction of the 
thermal neutrinos that radiate out with each neutrino depositing energy of the 
order of 10 MeV. Large convection in the central regions also helps this 
process.

\section{The neutrino oscillation probabilities}

The flavor eigenstate $|\nu_\alpha\rangle$ created inside the supernova
can be expressed as a linear superposition
of the mass eigenstates such that
$|\nu_\alpha\rangle = \sum_i U_{\alpha i} |\nu_i\rangle
$,
where $U$ is the unitary mixing matrix and the sum is over $N$ neutrino states.
After time t, the initial $|\nu_\alpha\rangle$ evolves to
$|\nu_\alpha(t)\rangle = \sum_i e^{-iE_i t}U_{\alpha i}|\nu_i\rangle$
where $E_i$ is the energy of the $i^{th}$
mass eigenstate. Then the probability of finding a flavor $\nu_\beta$ in the 
original $\nu_\alpha$ beam after traveling a distance L in vacuum 
is given by 
\be
P_{\alpha \beta} &=& \delta_{\alpha \beta} - 4~
\sum_{j > i}~ U_{\alpha i} U_{\beta i}^* U_{\alpha j}^* U_{\beta j}
\sin^{2}\left(1.27\frac{\Delta m^2_{ij} L}{E}\right)
\label{npr}
\ee
where $\Delta m^2_{ij} = m_{i}^2 - m_{j}^2$ is the mass squared difference.

Over the last few years the idea that neutrinos are not massless
but have small masses has become established
as a result of Super-Kamiokande (SK) and SNO which have firm evidence
for atmospheric and solar neutrino oscillations
\cite{Fukuda:1998mi,Fukuda:2002pe,Ahmad:2002jz,Ahmad:2002ka,
Bandyopadhyay:2002xj,Bahcall:2002hv}. The SK atmospheric neutrino 
data demand $\Delta m^2_{32} \sim 3\times 10^{-3}$ eV$^2$ and 
almost maximal mixing ($\sin^22\theta_{23} \approx 1$) 
\cite{Fukuda:1998mi} while the global solar neutrino data is best 
explained if $\Delta m^2_{21} \sim 6.1\times 10^{-5}$ eV$^2$ with 
large mixing angles ($\tan^2\theta \sim 0.41$) 
\cite{Bandyopadhyay:2002xj,Choubey:2002nc}. Very 
recently the KamLAND reactor antineutrino disappearance experiment 
\cite{kam_1} provided conclusive confirmation of the LMA solution to the 
solar neutrino problem, with mass and mixing parameters absolutely 
consistent with the solar neutrino results. The global analysis of the 
solar and the KamLAND data gives $\Delta m^2=7.17\times 10^{-5}$ eV$^2$ 
and $\tan^2\theta = 0.44$ \cite{Bandyopadhyay:2002en}. 
The CHOOZ reactor experiment restricts $\sin^2\theta_{13} <0.1$ 
for $\Delta m^2 > 10^{-3}$ eV$^2$ \cite{chooz}. 
The only other positive 
signal for neutrino oscillations come from the accelerator experiment 
LSND which requires $\Delta m^2\sim$ eV$^2$ and mixing angle 
small ($\sin^2 2\theta \sim 10^{-3}$).  To include LSND in 
the framework of oscillation one needs to extend the number of
neutrino generations to four, or in other words, include a sterile
neutrino. However with the latest SNO data on 
solar neutrinos and the final data from SK on atmospheric neutrinos, 
both the ``2+2'' and ``3+1'' 4-generation scenarios fail to explain
the global neutrino data. While ``3+1'' is inadequate in explaining
the combined accelerator-reactor data including LSND, ``2+2'' cannot
accomodate the solar and atmospheric neutrino data together 
\cite{Maltoni:2002ni}.

Since galactic supernova 
neutrinos with energies  $\sim 10$ MeV travel distances 
$\sim 10$ Kpc ($\sim 3\times 10^{20}$ m), the coherent 
term in eq.(\ref{npr}) becomes important only for $\Delta m^2_{ij}
\sim 10^{-19}$ eV$^2$. Thus supernova neutrinos can be used as probes for 
mass squared differences not possible to detect 
with any known terrestrial source.
However for the solar and atmospheric mass scales given above the oscillatory 
term would average out to 1/2.

The expression (\ref{npr}) 
would have been correct/exact if the neutrinos were
traveling in vacuum.
However for the supernova neutrinos things are a little complicated since 
they are created deep inside the core and traverse through extremely
dense matter before they come out into the vacuum. As the neutrinos 
move in matter they undergo scattering with the ambient electrons. 
While all the active neutrino flavors scatter electrons by the 
neutral current process, only the $\nue$ (and $\anue$) 
have charged current interactions as well. This 
significantly affects neutrino oscillations parameters as the 
$\nue$ picks up an additional matter induced mass term
\cite{Wolfenstein:1977ue}
\be
A(r)=2\sqrt{2} G_F N_An_e(r)E
\ee
where $G_F$ is the Fermi constant, $N_A$ the Avogadra's number, $n_e(r)$ 
the ambient electron density in the supernova at radius $r$ 
and $E$ the energy of the neutrino beam.
In appropriate units
\be
\frac{A(r)}{{\rm eV^2}}=
15.14\times 10^{-8} Y_e(r) \frac{\rho(r)}{{\rm gm/cm^3}}\frac{E}{{\rm MeV}}
\label{a}
\ee
where $Y_e(r)$ is the electron fraction and $\rho(r)$ is the 
matter density profile in the supernova which can be 
very well approximated by a power law $\rho(r)=Cr^{-n}$ with $n=3$ 
in the core \cite{bbb82}.

Neutrino oscillation probabilities may also be significantly affected 
inside Earth as the neutrinos traverse the Earth matter 
\cite{eartheffects}. 
Thus the neutrino oscillation probability is given by
\be
P_{\alpha\beta}=\sum_{i=1}^N P_{\alpha i}^{m}P_{i\beta}^{\oplus}
\label{p1}
\ee
where 
\be
P_{\alpha i}^{m} = \sum_{j=1}^N \left|U_{\alpha j}^m\right |^2 
\left|\langle \nu_i |\nu_j^m\rangle \right|^2
\label{p2}
\ee 
is the probability that a $\nu_\alpha$ ($\alpha=e,\mu,\tau$) 
produced inside the supernova core 
would emerge as a $\nu_i$ ($i=1,2,3$) at the surface of the supernova, 
$U_{\alpha j}^m$ are the elements of the mixing matrix 
at the point of production and 
$|\langle \nu_i|\nu_j^m\rangle |^2$ is the probability that 
a $|\nu_j^m\rangle$ state in matter appears as the state 
$|\nu_i\rangle$ at the supernova surface in vacuum. This is the so called 
``jump probability''.
$P_{i\beta}^{\oplus}$ is the 
probability that the $\nu_i$ mass 
eigenstate arriving at the surface of the Earth 
is detected as a $\nu_\beta$ flavor state in the detector. Depending on 
whether the neutrinos cross the Earth or not, 
$P_{i\beta}^{\oplus}$ maybe different 
from $|U_{\beta i}|^2$, where $U_{\beta i}$ is the element of the
mixing matrix in vacuum.

Since to a good approximation the average energy and the total 
fluxes of $\numu$, $\anumu$, $\nutau$ and $\anutau$ are same, 
for mixing between only active neutrino flavors the 
only relevant oscillation probability that we need is the $\nue$ survival 
probability $P_{ee}$ which is given by eq.(\ref{p1}) with $\alpha=\beta=e$.

\subsection{Two flavor oscillations}

To begin with let us for simplicity assume that there are just two 
neutrino flavors, $\nue$ and another active flavor $\nu_a$ which may be
$\numu$ or $\nutau$.   
The effective mixing angle in matter is given by 
\be
\tan2\theta_m(r) = \frac{\Delta m^2 \sin2\theta}
{\Delta m^2\cos2\theta - A(r)}
\label{mix2}
\ee
Since the density inside the supernova core where the neutrinos 
are created is extremely high, $A(r_s) \gg \Delta m^2$ and 
$\theta_m \approx \pi/2$. Hence the 
survival probability $P_{ee}$ given by eq.(\ref{p1}) and (\ref{p2}) 
reduces to
\be
P_{ee}=P_JP_{1e}^\oplus + (1-P_J)P_{2e}^\oplus
\label{p2gen}
\ee
where $P_J=|\langle \nu_1|\nu_2^m\rangle|^2$ 
is the jump or the crossing probability 
from one neutrino mass eigenstate to the other at resonance. 
If $P_J\approx 0$ the neutrino propagation in matter 
is called {\it adiabatic}, otherwise its
{\it non-adiabatic}. When $P_J \rightarrow 1$ then we encounter the extreme
non-adiabatic situation. In \cite{Fogli:2001pm} it is shown that
the {\it double-exponential} parametrization of $P_J$ 
derived in \cite{Petcov:1987zj} and used 
extensively for the solar neutrinos, works 
extremely well even for the supernova density profile. In this 
parametrization the jump probability is expressed as
\be
P_J = \frac{\exp(-\gamma \sin^2 \theta) - \exp(-\gamma)}{1-\exp(-\gamma)}
\label{pjump}
\ee
where $\gamma$ is given by
\be
\gamma =\pi\frac{\Delta m^2}{E}
\left|\frac{dln~n_e}{dr}\right|_{r=r_{\it mva}}^{-1}
\label{gamma}
\ee
The {\it density scale factor} 
$\left|\frac{dln~n_e}{dr}\right|^{-1}$ gives a measure of 
the deviation from adiabaticity and is calculated at the position where 
we have {\it maximum violation of adiabaticity (mva)} 
\cite{Fogli:2001pm,Lisi:2000su}. That is where
\be
A(r_{\it mva})=\dm
\label{rmva}
\ee
Note that the position of {\it mva} ($r_{mva}$) 
is different from the 
position of resonance ($r_{res}$) which is given by the condition
\be
A(r_{res})=\Delta m^2\cos2\theta
\label{res}
\ee

The form of the probability $P_{ie}^\oplus$ depends crucially 
on the trajectory of the neutrinos inside the earth and hence on the 
direction of the supernova. If the direction is such that the neutrinos 
cross only the mantle of the Earth then the amplitude
\be
A_{2e}^\oplus&=&\sum_j U_{ej}^e e^{-i\phi_j^e}\langle
\nu_j^e|\nu_i\rangle
\ee
where $U_{ej}^e$ is the mixing matrix elements in the Earth's mantle
and $\phi_j^e$ is the phase. Therefore the expression for 
$P_{2e}^\oplus(=1-P_{1e}^\oplus)$ is given by
\be
P_{2e}^\oplus=\sin^2\theta + \sin2\theta_e\sin(2\theta_e-2\theta)
\sin^2\left(1.27 \frac{\Delta m_e L}{E}\right )
\label{oneslab1}
\ee
where $L$ is the distance traversed inside Earth and 
$\theta_e$ (given by eq.(\ref{mix2}) but with $A$ calculated 
in the mantle of the Earth) and $\Delta m^2_e$ are the mixing 
angle and the mass squared 
difference inside the Earth's mantle.  
If the neutrinos cross both the mantle as well as the core of the Earth then
\be
A_{2e}^{\oplus}&=&
\sum_{\stackrel{i,j,k,}{\alpha,\beta,\sigma}}
U_{ek}^M e^{-i\psi_k^M}
U_{\alpha k}^M U_{\alpha i}^C
e^{-i\psi_i^C}
U_{\beta i}^C U_{\beta j}^M
e^{-i\psi_j^M}
U_{\sigma j}^M U_{\sigma 2}
\ee
where ($i,j,k$) denotes mass eigenstates and
($\alpha,\beta,\sigma$) denotes flavor eigenstates, $U$,
$U^M$ and $U^C$ are the mixing matrices in vacuum, in the mantle and the
core respectively and $\psi^M$ and $\psi^C$ are the corresponding phases
picked up by the neutrinos as they travel through the mantle and the core
of the Earth. Then the probability
\be
P_{2e}^\oplus= |A_{2e}^\oplus|^2
\label{twoslab1}
\ee

The additional mass term picked up by the $\anue$ as it moves 
in matter is $-A(r)$. Since the crucial combination which decides 
matter effects is the ratio $A(r)/\Delta m^2$,  
the antineutrino survival probability $\bar P_{ee}$ is identical 
to that for the neutrinos if we change the sign of $\Delta m^2$, 
which is equivalent to swaping of the mass labels $1\leftrightarrow 2$ 
\cite{Fogli:2001pm}. 
Then the 
expression for  $\bar P_{ee}$ is similar to that for $P_{ee}$ 
and is given by
\be
\bar P_{ee}=\bar P_J \bar P_{1e}^\oplus + (1-\bar P_J)\bar P_{2e}^\oplus
\label{p2gena}
\ee
where 
\be
\bar P_J = \frac{\exp(-\gamma \cos^2 \theta) - \exp(-\gamma)}{1-\exp(-\gamma)}
\label{pjumpa}
\ee
where we replace $\cos^2\theta$ with $\sin^2\theta$ (swaping 
 $1\leftrightarrow 2$)
and $\gamma$ is calculated at $r_{mva}$ given by the same eq.(\ref{rmva}).
The expressions for the oscillation probabilities $\bar P_{2e}$ are again 
similar to those for the neutrinos
\be
\bar P_{2e}^\oplus&=&\sin^2\theta + \sin2\bar\theta_e\sin(2\bar\theta_e-2\theta)
\sin^2\left(1.27 \frac{\bar{\Delta m_e} L}{E}\right )
\label{oneslab1a}
\ee
where Eq.(\ref{oneslab1a}) is for transition probability in Earth for 
one slab approximation, with the mixing angle 
$\bar\theta_e$ given by
\be
\tan2\bar\theta_e = \frac{\Delta m^2 \sin2\theta}
{\Delta m^2\cos2\theta + A(r)}
\label{mix2a}
\ee
The expression for $\bar P_{2e}^\oplus$ for two slabs can also be
similarly derived from 
(\ref{twoslab1}) with the corresponding changes for the antineutrinos.

\subsection{Three flavor oscillations}

We now consider a more realistic scenario with mixing between 
three active neutrinos, with one of the mass squared difference 
corresponding to the solar scale ($\Delta m^2_{21} \sim 10^{-5}$ eV$^2$) 
and the other one corresponding to the atmospheric scale 
($\Delta m^2_{31}\sim 10^{-3}$ eV$^2$). In this case the 
neutrinos encounter two resonances, the first one corresponding 
to the higher scale at a higher density in the supernova
and the next one corresponding 
to the lower mass scale further out in the mantle.
Though from solar neutrino
data we know that the sign of $\Delta m^2_{21}\equiv \Delta m^2_\odot$
is positive
\cite{Bandyopadhyay:2002xj}
($\Delta m_{ij}^2=m^2_i-m^2_j$), there is still an ambiguity
in the sign of $\Delta m^2_{32}\equiv\Delta m^2_{atm}$.
It would be hard to determine the sign of $\Delta m^2_{32}$ in any of
the current and planned long baseline oscillation experiments and
only a neutrino factory would be able to resolve this ambiguity.
However for the supernova neutrinos the sign of 
$\Delta m^2_{32} \sim \Delta m^2_{31}$ is crucial and thus 
the supernova neutrinos can be used very effectively 
to give us the neutrino mass 
hierarchy.

\subsubsection{Direct mass hierarchy}

Since the density 
at the neutrino source ($r_s$) is very high,
$A(r_s) \gg  \Delta m^2_{31} \gg \Delta m^2_{21}$ and  
we can solve the  eigenvalue problem perturbatively to get 
the mixing angles for neutrinos in matter\footnote{If we choose the standard 
parametrization of the mixing matrix, the mixing angle 
$\theta_{23}$ does not affect the $\nue$ survival probability and 
thus we can either choose to rotate it away or even put it 
to zero without loss of generality.} 
\cite{Fogli:2001pm,Kuo:1989qe,Kuo:1987qu,Dutta:1999ir}
\be
\tan2\theta_{12}^m(r) &=& \frac{\Delta m^2_{21}\sin2\theta_{12}}
{\Delta  m^2_{21}\cos2\theta_{12} - \cos^2\theta_{13}A(r)}
\label{theta12m}
\\
\tan2\theta_{13}^m(r) &=& \frac{\Delta m^2_{31}\sin2\theta_{13}}
{\Delta  m^2_{31}\cos2\theta_{13} - A(r)}
\label{theta13m}
\ee
At the point of production since 
$A(r_s)\gg  \Delta m^2_{31} \gg \Delta m^2_{21}$ from 
eq.(\ref{theta12m}) and (\ref{theta13m}) we see that 
$\theta_{12}^m \approx \pi/2 \approx \theta_{13}^m$ and neutrinos 
are created in almost pure $\nu_3^m$ states and 
the expression for the survival probability for this three-generation 
scenario is
\be
P_{ee}=P_HP_LP_{1e}^\oplus +P_H(1-P_L)P_{2e}^\oplus +(1-P_H)P_{3e}^\oplus
\label{prob3}
\ee
where $P_H$ and $P_L$ are the jump probabilities for the high and low 
density transitions respectively. Just like in the two-generation case 
they can be calculated using the double exponential forms with
\be
P_{L} &=& \frac{\exp(-\gamma_{L} \sin^2 \theta_{12}) - 
\exp(-\gamma_{L})}{1-\exp(-\gamma_{L})}
\label{pjumpL}
\\
P_{H} &=& \frac{\exp(-\gamma_{H} \sin^2 \theta_{13}) - 
\exp(-\gamma_{H})}{1-\exp(-\gamma_{H})}
\label{pjumpH}
\ee
where $\gamma_{L,H}$ is calculated using eq.(\ref{gamma}) 
at the position of maximum violation of adiabaticity 
corresponding to the lower ($r_L$) and the higher scales 
($r_H$) respectively given 
by the relations
\be
\cos^2\theta_{13}A(r_{L})=\Delta m^2_{{21}}
\label{gammaL3}
\\
A(r_{H})=\Delta m^2_{{31}}
\label{gammaH3}
\ee

For the antineutrinos $\anue$ the matter term is negative so as in 
the two-generation case
the mixing angle for the antineutrinos in matter is given by 
\be
\tan2\bar\theta_{12}^m(r) &=& \frac{\Delta m^2_{21}\sin2\theta_{12}}
{\Delta  m^2_{21}\cos2\theta_{12} + \cos^2\theta_{13}A(r)}
\label{atheta12m}
\\
\tan2\bar\theta_{13}^m(r) &=& \frac{\Delta m^2_{31}\sin2\theta_{13}}
{\Delta  m^2_{31}\cos2\theta_{13} + A(r)}
\label{atheta13m}
\ee
which implies that at the point of production 
$\cos2\bar\theta_{12}^m \approx +1 \approx \cos2\bar\theta_{13}^m$
($\bar\theta_{12}^m \approx 0 \approx \bar\theta_{13}^m$) and the 
antineutrinos are created in pure $\bar\nu_1^m$ state. Thus for the 
antineutrinos the survival probability is given by \cite{Fogli:2001pm}
\be
\bar P_{ee} = (1-\bar P_L)\bar P_{1e}^\oplus + \bar P_L \bar P_{2e}^\oplus
\label{aprob3}
\ee
where the jump probability $\bar P_L$ for the antineutrinos is given by
\be
\bar P_L=\frac{\exp(-\gamma_{L} \cos^2 \theta_{12}) - 
\exp(-\gamma_{L})}{1-\exp(-\gamma_{L})}
\label{apjumpL}
\ee
with $\gamma_L$ defined by eq.(\ref{gammaL3}) and (\ref{gamma}).

\subsubsection{Inverse mass hierarchy}

If $\Delta m^2_{31}\approx \Delta m^2_{32} $ 
is negative, the mixing angles for the neutrinos are 
still given by the eq.(\ref{theta12m}) and (\ref{theta13m}) but with 
the sign of $\Delta m^2_{31}$ reversed\footnote{Note that the we take the 
sign of $\Delta m^2_{21}$ as positive in accordance with the 
currently favored LMA MSW solutions 
to the solar neutrino problem \cite{Choubey:2002nc}.}. At the production point 
then $\theta_{12}^m \approx \pi/2$ while $\theta_{13}^m \approx 0$ and 
$\nue$ are thus in almost 
pure $\nu_2^m$ states and the neutrino survival probability is
\be
P_{ee} = P_L P_{1e}^\oplus + (1-P_L)P_{2e}^\oplus
\label{invprob3}
\ee
with the jump probability $P_L$ given by eq.(\ref{pjumpL}), 
(\ref{gammaL3}) and (\ref{gamma}).

With inverse hierarchy the antineutrino mixing angles are given by 
eq.(\ref{atheta12m}) and (\ref{atheta13m}) with the sign of 
$\Delta m^2_{31}$ reversed. Therefore $\anue$ are created in pure 
$\bar \nu_3^m$ states and their survival probability is \cite{Fogli:2001pm}
\be
\bar P_{ee} = (1-\bar P_L)P_H\bar P_{1e}^\oplus + 
\bar P_L P_H \bar P_{2e}^\oplus + (1-P_H)\bar P_{3e}^\oplus
\label{ainvprob3}
\ee
with $\bar P_L$ given by eq.(\ref{apjumpL}) and $P_H$ by eq.(\ref{pjumpH}).

\section{Event rates in terrestrial detectors}


Neutrinos are created deep inside the supernova core as $\nu-\bar\nu$ 
pairs. They stream out through the supernova core, mantle and envelope 
and reach the Earth after travelling distances $\sim 10^{17}$ km.  
In presence of neutrino 
oscillations there is a modification of the neutrino fluxes 
as they oscillate into one another and the resultant neutrino beam at Earth is
given by
\be
N_{\nu_e} &=& P_{ee}(E)N_{\nue}^0(t)+P_{\mu e}(E)N_{\numu}^0(t)
+P_{\tau e}(E)N_\nutau^0(t)\nonumber\\
&=&P_{ee}(E)N_{\nue}^0(t)+(1-P_{ee}(E))N_{\nu_x}^0(t)
\label{fluxe}\\
N_{\anue} &=&\bar P_{ee}(E)N_{\anue}^0(t)+(1-\bar P_{ee}
(E))N_{\nu_x}^0(t)
\label{fluxae}\\
N_{\nu_x} &=&(1-P_{ee}(E))N_{\nue}^0(t)+(1+P_{ee}(E))N_{\nu_x}^0(t)
\label{fluxx}\\
N_{\bar\nu_x} &=&(1-\bar P_{ee}(E))
N_{\nue}^0(t)+(1+\bar P_{ee}(E))N_{\nu_x}^0(t)
\label{fluxax}
\ee
where $P_{ee}$ and $\bar P_{ee}$ are the $\nue$ and $\anue$
survival probabilities given in the previous section and 
$N_{\nu_\alpha}^0(t)$ is the neutrino flux produced inside the
supernova core given by 
$N_{\nu_\alpha}^0(t)= L_{\nu_\alpha}(t)/\langle E_{\nu_\alpha}(t)\rangle$,
where $L_{\nu_\alpha}(t)$ is the neutrino luminosity and
$\langle E_{\nu_\alpha}(t) \rangle$ is the average energy. 
In the above expressions we have used the fact that the $\numu/\anumu$ beam is
indistinguishable from the $\nutau/\anutau$ beam in flux and energy 
and call them $\nu_x$.  

The current and planned terrestrial detectors are capable of 
observing the supernova neutrinos through various charged and
neutral current processes.
The differential number of neutrino events at the detector for a
given reaction process is
\begin{equation}
\frac{d^2 S}{dE_\nu dt}=\sum_\alpha\frac{n}{4\pi D^2}
N_{\nu_\alpha}f_{\nu_\alpha}(E_\nu) \sigma(E_\nu)
\epsilon(E_\nu)
\label{sig}
\end{equation}
where $\alpha$ runs over the neutrino species concerned (e, $\mu$, $\tau$),
$N_{\nu_\alpha}$
is the neutrino flux {\it at the detector} given by eqs. 
(\ref{fluxe})--(\ref{fluxax}) 
and $\sigma (E_\nu)$ is the reaction cross-section for the neutrino with
the target particle, $D$ is the distance of the neutrino source
from the detector (taken as 10kpc for galactic supernovea considered here),
$n$ is the number of detector particles for the reaction considered
and $f_{\nu_\alpha} (E_\nu)$ is the energy spectrum for the neutrino species
involved,
while $\epsilon(E_\nu)$ is the detector efficiency as a function of the
neutrino energy. 

The main reaction process by which the water \chr detectors like SK 
would observe the supernova neutrinos is
\be
\anue + p \rightarrow e^+ + n
\label{anuep}
\ee
However the other neutrino species are also observable in SK through the 
$\nu-e$ elastic scattering processes
\be
\nu_i + e^- \rightarrow \nu_i + e^-
\label{elas}
\ee
In addition to the above two reactions, supernova neutrinos can also 
be traced in the water 
\chr 
detectors 
through reactions involving $^{16}O$. The oxygen nuclei 
in water are doubly closed shell and have a very high threshold
($E_{th}$) for
excitation. Thus solar neutrinos are unable to have charged or neutral 
current reactions on oxygen. But supernova neutrinos with much larger
energy range can trigger charged current reactions \cite{Haxton:kc}
\be
\nue + ^{16}O &\rightarrow& e^- + ^{16}F
\label{oxycce}\\
\anue + ^{16}O &\rightarrow& e^+ + ^{16}N
\label{oxyccae}
\ee
and neutral current reaction \cite{Langanke:1995he}
\be
\nu_x + ^{16}O \rightarrow \nu_x + ^{16}O^*
\label{oxync}
\ee
where $^{16}O^*$ decays by $n$, $p$ or $\gamma$ emission. The 
reaction thresholds for the charged current reactions (\ref{oxycce}) 
and (\ref{oxyccae}) are 15.4 MeV and 11.4 MeV respectively 
\cite{Haxton:kc}.
The electrons from the charged current reactions on $^{16}O$ can be 
distinguished in principle from the positrons from $\anue$ capture 
on protons (cf. reaction (\ref{anuep}))by their
angular distribution. While the $^{16}O$ events are backward peaked 
and electron scattering events are strongly forward peaked, the 
$\anue p$ events are mostly isotropic. Thus even though all these
processes are detected via the \chr tecnique, its possible to
disentangle them.

In heavy water ($D_2O$) detectors like SNO, in addition to the reactions 
involving elastic scattering off electrons and reactions on $^{16}O$, 
neutrinos can be observed by the charged and neutral current 
breakup of deuteron
\be
\nue + d &\rightarrow& p+p + e^-
\label{nued}
\\
\anue + d &\rightarrow& n+n+e^+
\label{anued}
\\
\nu_x + d &\rightarrow& p+n+\nu_x
\label{nuxd}
\ee
The charged current reactions are detected by the \chr radiation from
the electron/positron. The neutral current reaction, which will give
us information about the total neutrino flux from the supernova,
irrespective of whether they oscillate or not, is detected by the
capture of the released neutron, either on deuteron or on $^{35}Cl$ (salt). 
In the last phase of SNO the neutral current process will be detected
by directly observing the neutrons in helium proportional counters.

There have been various attempts before to estimate the effect of
non-zero neutrino mass and mixing on the expected neutrino signal
from a galactic supernova.
With vacuum
oscillations we can expect an increase in both the $\nue$ and $\anue$
signal \cite{bkg26,Choubey:1998nh}. Some special cases where the 
matter effects inside the supernova 
are negligible and one has almost pure vacuum
oscillations have been considered in \cite{Choubey:1998nh}. 
However for the currently most prefered neutrino mass spectrum one 
expects to have substantial matter effects.
Matter enhanced resonant flavor conversion has been observed to have
a large effect on the $\nue$ signal 
\cite{bkg26,kp6,qf6,ds6,Dutta:2001nf,Dutta:2000zq,akh6}.

\begin{table}[t]
    \begin{center}
\begin{tabular}{||c| c| c|c||} \hline\hline
&&A&B\\\hline
reactions&
{$\nu_e+d\rightarrow p+p+e^-$} & { 75 } & {239} \\
in&
{$\bar\nu_e +d\rightarrow n+n+e^+$} & {91} & {91} \\
1 kton&
{$\nu_i+d\rightarrow n+p+\nu_i$} & {544} & {544} \\
$\rm{D_2O}$&
{$\nu_e +e^- \rightarrow \nu_e +e^-$} & {4} & {6} \\
&
{$\bar\nu_e+e^- \rightarrow \bar\nu_e+e^-$}
& {1} & {1} \\
&
{$\nu_{\mu,\tau}(\bar\nu_{\mu,\tau}) +e^- \rightarrow
\nu_{\mu,\tau}(\bar\nu_{\mu,\tau}) +e^-$}
& {4} & {3} \\
&
{$\nu_e +^{16}O \rightarrow e^- +^{16}F$} & {1} & {55} \\
&
{$\bar\nu_e + ^{16}O\rightarrow e^+ + ^{16}N$} & {4} & {4} \\
&
{$\nu_i +^{16}O \rightarrow \nu_i +\gamma +X$} & {21} & {21} \\ \hline

reactions&
{$\bar\nu_e +p\rightarrow n+e^+$} & {357} & {357} \\
in &
{$\nu_e +e^- \rightarrow \nu_e +e^-$} & {6} & {9} \\
1.4 kton&
{$\bar\nu_e+e^- \rightarrow \bar\nu_e+e^-$} & {2} & {2} \\
$\rm{H_2O}$ &
{$\nu_{\mu,\tau}(\bar\nu_{\mu,\tau}) +e^- \rightarrow
\nu_{\mu,\tau}(\bar\nu_{\mu,\tau}) +e^-$} & {6} & {5} \\
&
{$\nu_e +^{16}O \rightarrow e^- +^{16}F$} & {2} & {86} \\
&
{$\bar\nu_e + ^{16}O\rightarrow e^+ + ^{16}N$} & {6} & {6} \\
&
{$\nu_i +^{16}O \rightarrow \nu_i +\gamma +X$} & {33} & {33} \\
\hline\hline
\end{tabular}
\end{center}
\caption[Signal from a galactic supernova for complete conversion]
{\label{mswtab}
The expected number of neutrino events in SNO. To get
the number of events in SK, one has to scale the number of events in
$\rm{H_2O}$ given here to its fiducial mass of 32 kton.
The column A corresponds to massless neutrinos,
column B to neutrinos with complete flavor conversion ($P_{ee}\approx 0$).
The mixing angle $\theta_{12}$ is considered to be very small 
corresponding to the SMA solution and hence $\bar P_{ee}\approx 1$.
The $\nu_i$ here refers to all the six neutrino species.}
\end{table}

Table \ref{mswtab}
gives the calculated number of expected events for the
main reactions in $\rm H_2O$ and $\rm D_2O$, for a typical 
galactic supernova with a total luminosity of about $3\times 10^{53}$ ergs.  
The numbers here correspond to a three-flavor 
oscillation scenario with complete flavor conversion.
The $\theta_{13}$ considered here is {\it large} so that  
both $P_L$ and $P_H$ are almost zero, the propagation is almost 
adiabatic and hence $P_{ee}\approx 0$. 
The $\theta_{12}$ considered is very small and hence  
$\bar P_{ee} \approx 1$\footnote{The Table \ref{mswtab} is just for the 
purpose of illustration only. For the LMA solution the $\nue$ events
would still remain the same, while the $\anue$ events would be
slightly enhanced.}.
For the cross-section of the $(\nue-d), (\anue-d),
(\nu_x-d)$
and $(\anue-p)$ reactions we refer to \cite{Burrows:1990}.
The cross-section of the $(\nue(\anue)-e^-)$ and $(\nu_x-e^-)$
scattering has been taken from \cite{kolb6} while the neutral current
$(\nu_x-^{16}O)$ scattering cross-section is taken from \cite{bv16}.
For the 
$^{16}O(\nue,e^-)^{16}F$ and $^{16}O(\anue,e^+)^{16}N$ reactions we
refer to \cite{Haxton:kc} where we have used the cross-sections
for the detector with perfect efficiency.

From a comparison of the
predicted numbers in Table \ref{mswtab},
it is evident that neutrino oscillations
play a significant role in supernova neutrino detection.
As the average energy of the $\numu/\nutau$ is greater
than the average energy of the $\nue$, neutrino flavor mixing
modifies their energy spectrum.
Hence 
the $\nue$ flux though depleted in number, gets enriched
in high energy neutrinos and since the detection
cross-sections are strongly energy dependent, this results in the
enhancement of the charged current signal \cite{Choubey:1998nh}. Since the 
cross-section for the $^{16}O$ reactions have the strongest 
dependence on energy, they are most affected by neutrino 
oscillations and can be used as an effective way to study neutrino 
properties from supernova neutrino detection.
For the neutral current sector the number of
events remain unchanged as the interaction is flavor blind.

\begin{figure}
\vskip -1truecm
\centerline{\epsfig{file=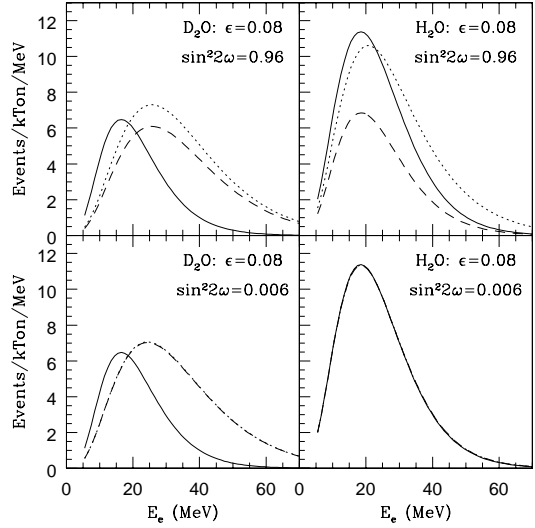,width=4.0in}}
\vskip -0.5in
\caption{Total event rates (from combining the indivdual $\nu_e\, d$
and $\overline{\nu}_e\, d$) are shown as a function of the electron/positron
energy, $E_e$, for two different values of $\omega\equiv\theta_{12}$, 
and for $\epsilon\equiv \sin^2\theta_{13}
= 0.08$ so that the propagation is fully adiabatic. The dotted and
dashed lines are due to the effects of 3- and 4-flavour mixing.
Results from a 1 kton water detector (from $\overline{\nu}_e\, p$
alone) are shown on the right, for comparison. This figure is taken
from \cite{Dutta:2001nf}.}
\label{murthy1} 
\end{figure}

\begin{figure}
\vskip -1cm
\centerline{\epsfig{file=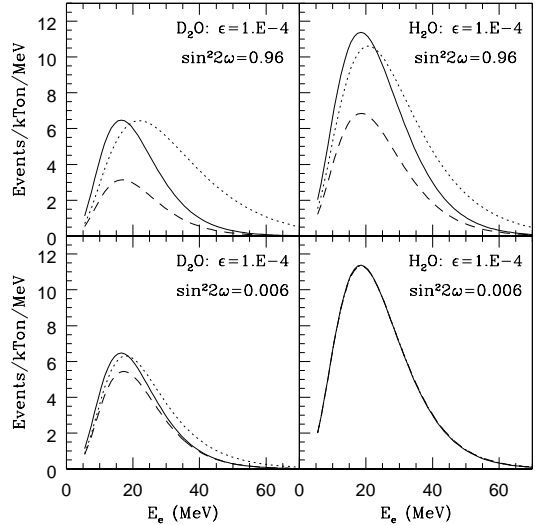,width=4.0in}}
\vskip -0.5in
\caption{Same as Fig.~\ref{murthy1} but for $\epsilon \sim 0$ so that
non-adiabatic effects are included. 
This figure is taken
from \cite{Dutta:2001nf}.}
\label{murthy2} 

\end{figure}

Figure \ref{murthy1} taken from \cite{Dutta:2001nf}, shows the 
comparison between the total charged current events as a
function of the electron/positron energy  
observed in $H_2O$ ($\anue p$ events) and $D_2O$ (sum of $\nue d$ and $\anue
d$ events) for small and large values of 
the mixing angle $\sin^22\theta_{12}$ ($\omega \equiv \theta_{12}$). 
The value of $\sin^2\theta_{13}\equiv \epsilon$ is large ($=0.08$) 
which implies that the neutrino propagation is fully adiabatic. 
Figure \ref{murthy2} also taken from \cite{Dutta:2001nf}, shows the 
corresponding plots when $\sin^2\theta_{13}\equiv \epsilon$ is small 
($\sim 0$), which implies that the neutrino propagation is
non-adiabatic.
In both the figures the solid lines give the no oscillation
distribution and the dotted line the events for three-generation
scenario, while the dashed lines correspond to the distribution for 
a four-generation scheme\footnote{We have not considered the
four-generation scenario in this review. For a detailed discussion 
on the four-generation neutrino mass spectrum and its effects 
on supernova neutrino detection refer to
\cite{Dutta:2000zq,Dutta:2001nf}.}.
The figures show that for the LMA solution (upper panels) there is a 
shift in the spectral peak in $D_2O$ for both three as 
well as four generations,
the shift being more pronounced for the high $\epsilon$ (adiabatic)
case. The corresponding shifts in $H_2O$ are less. 
The SMA cases shown in the lower panels are of less interest now 
since SMA solution is now ruled out. The figures show that by
comparing the signal in SK and SNO one can distinguish between the 
three and four-generation scenario. But again the four-generation
schemes are largely disfavored by the global solar and atmospheric
neutrino data \cite{Maltoni:2002ni}.

The potential for detecting supernova neutrinos in scintillation
detectors like Borexino \cite{Cadonati:2000kq} and MiniBOONE
\cite{Sharp:2002as} have been recently considered. The 
$^{12}C$ in these detectors can be excited through charged current and 
neutral current interations with the supernova neutrinos. The charged
current reaction ($^{12}C(\nue,e^-)^{12}N$) 
has a threshold of 17.34 MeV while that for ($^{12}C(\anue,e^+)^{12}B$) 
has a threshold of 14.4 MeV.   
But here again the cross-sections have a strong energy dependence 
and these events show a dramatic increase with large conversion with
oscillations compared to the no oscillation case. The neutral current 
events through ($^{12}C(\nu_x,\nu_x)^{12}C^*$) can be used to put
direct limits on the neutrino masses using the time delay techniques
briefly discussed in the following section.

\section{Other effects of neutrino mass and mixing}
In this section we briefly touch upon a number of areas where the mass and 
mixing of supernova neutrinos can lead to interesting effects:

a). {\bf SN 1987A:} The eleven SN 1987A events at KII were observed within a
timespan of 5.6 secs and with an energy range of the positron/electron released
in the water Cerenkov detector 
from 7.5 MeV to 35.4 MeV. Similarly IMB had the 
eight events within a time of 12.4 secs and with the electron energy range of
20 MeV to 40 MeV \cite {Schramm:1990}. The angle of the $e^{+}/e^{-}$ path to
the $\nu/\bar{\nu}$ direction for each event was also measured. There were also
5 events at Mt. Blanc and 3 in Baksan at the same time \cite {Schramm:1990}.
A number of analysis were done in the next few years and the results 
more-or-less agreed with the typical values given in section 2 for the 
luminosities, average energies and spectra of the neutrinos, though the IMB
events gave average energy and temperature consistently higher
\cite {Burrows:1987,Kahana:1987}. Also with such
small samples there were large errors in the extraction of the SN parameters.
However even though the statistics were poor the SN1987A data was 
used extensively to study the neutrino mass and mixing patterns.
In the context of two flavors such analysis was done by Smirnov et al
\cite {Smirnov:1994} and Jegerlehner et al \cite {Jeger:1996} and
recently it
was extended to three flavors in \cite {Minakata:2000rx}.
The authors of \cite{Minakata:2000rx} claim that the inverse mass hierarchy is
disfavored by the data unless $\theta_{13}$ is very small,
$\sin^2\theta_{13} <10^{-4}$. However the authors of
\cite{Barger:2002px} dispute this observation and conclude 
that the SN1987A data cannot distinguish between the direct and 
inverted mass hierarchies. In \cite{Kachelriess:2000fe} the SN1987A
data is combined with the global solar neutrino data and it is found 
that while all the other large mixing angle solutions (LOW-QVO and VO) 
are disfavored, 
the LMA solution remains the only allowed solution which can explain
the SN1987A  and the solar neutrino observations simultaneously.
Nowadays after the evidence of neutrino mass and mixing one has to work on the
``inverse problem'' using SN 1987A data to extract the original neutrino spectra
using realistic (Large Mixing Angle solution) scenario of neutrino
oscillation \cite{Minakata:2001cd,Barger:2001yx}.

a). {\bf Detection of neutronisation neutrinos}: The neutrinos emitted during
the collapse phase due to the neutronisation give rise to a luminosity small
compared to the thermal post-bounce neutrinos discussed above, but for close
enough (1 kpc) galactic supernovae they can still be detected by SK and SNO
\cite {Sutaria:1997}. The measurement of the fluence of these neutrinos at SNO
and the distortion of the spectrum detected at SK, in particular the ratio of
the calorimetric detection of the neutrino flunce via the neutral current
channel to the total energy integrated fluence observed via the charged
current channel at SNO can yield valuable infirmation about the mass squared
difference and mixing \cite{Majumdar:1998vf}.

b). {\bf Delay of massive supernova neutrinos}: For a neutrino of mass {\bf m}
(in eV) and energy {\bf E} (in MeV) the delay (in sec) in traveling a distance
(in 10 kpc) is
\be
        \Delta t(E)=0.515 (m/E)^2 D
\ee
neglecting small higher order terms. If we assume that the mass of the 
$\nu_x$ is much larger than those of $\nue$ and $\anue$ then the
neutral current events will have a delay compared to the charged
current events. This difference due to 
time-of-flight for neutral current 
signal compared to the charged current signal in SNO can determine $\nu_{\mu}$
and $\nu_{\tau}$ mass down to 30 eV, an improvement by many orders of
magnitude over current estimates \cite{bv16}.
One also sees that one can construct useful diagnostic tools for neutrino
mass and mixing using the charged and neutral current events as a function of 
time but only for mass squared differences of the order of tens of 
$eV^2$ \cite{Choubey:2000ma}.

c). {\bf Effect of neutrino mixing on delayed neutrino heating}: To generate
a stronger shock in the supernova models one thinks of mechanisms of extra
heating in the region near the shock. As the heating rate due to neutrino 
capture depends on the square of the neutrino temperature, if the $\nu_{\mu}$
or $\nu_{\tau}$ emitted from the neutrino sphere 
can get converted to $\nu_e$ before reaching the shockfront, 
it heats up the shock more.
Fuller et al \cite{Fuller:1990} in their numerical calculations got 60\%
more heating but with the $\nu/\tau$ neutrino mass of 40 eV. However 
with realistic
solar and atmospheric mass squared differences one does not get this conversion
to $\nu_e$ inside the stalled shock.
Recently it is proposed  that the neutrino signal in present and future
neutrino detectors can give valuable information about the mechanism of shock propagation
 and the delayed neutrino heating \cite{Schirato:2002tg}. When the shock
front moves through the MSW conversion region the $\mu,\tau$ to $e$ type
neutrino conversion gets stopped during that time leading to a detectable
dip in the neutrino energy/count rate.

d). {\bf r-process nucleosynthesis:} The 
neutrino-driven-wind environment in the
late time (about 3--15 secs after bounce) 
of core collapse supernova is considered
to be a very promising site for the rapid neutron capture process (r - process)
for producing neutron-rich heavy elements. The capture rate of $\nue$ and
$\anue$ on neutrons and protons respectively determine the electron fraction,
$Y_e$ and for successful r-process $Y_e$ must be less than 0.5. This is favored
by the higher average energy of $\anue$; however if oscillations between 
$\nue$ and
$\nu_x$ takes place giving a stiffer $\nue$ spectrum, the r-process may get
stopped. Thus to get r-process nucleosynthesis operative one excludes
the parameter space   
$\Delta m^2 >$ a few $eV^2$ and $\sin^2 2\theta < 10^{-5}$ \cite{Qian:1993}.
Recently 
the effect of active-sterile neutrino transformation
on the r-process was also considered in \cite{Fuller:2002} 
and initial work showed that it is possible to get
sufficiently neutron rich matter to activate rapid neutron capture.

\begin{flushleft}
{\bf Acknowledgement}
\end{flushleft}
We thank G.Dutta, D. Indumathi, M.V.N. Murthy and
G. Rajasekaran for allowing us to reproduce two of their figures.

\end{document}